\documentclass[twoside,10pt]{article}
\usepackage{epsfig}
\usepackage{amsmath,amsfonts}

\newcommand{\be}{\begin{equation}}
\newcommand{\ee}{\end{equation}}
\newcommand{\bea}{\begin{eqnarray}}
\newcommand{\eea}{\end{eqnarray}}

\begin{document}

\title{  Constraints on the inner edge of neutron star crusts from
relativistic nuclear energy density functionals
}

\author{Ch.C. Moustakidis$^{1}$,  T. Nik\v si\'c$^{2}$, G.A. Lalazissis$^{1}$, D. Vretenar$^{2,3}$,
 and P. Ring$^{3}$ \\
\\
$^{1}$Department of Theoretical Physics, Aristotle University of
Thessaloniki, \\ 54124 Thessaloniki, Greece \\
$^{2}$Physics Department, Faculty of Science, University of
Zagreb,
\\ 10000 Zagreb, Croatia\\
$^{3}$Physik-Department der Technischen Universit\"at M\"unchen, \\
D-85748 Garching, Germany\\
}

\maketitle
\begin{abstract}
The transition density $n_t$ and pressure $P_t$ at the inner edge between
the liquid core and the solid crust of a neutron star are analyzed using the
thermodynamical method and the framework of relativistic nuclear energy
density functionals. Starting from a functional that has been carefully adjusted
to experimental binding energies of finite nuclei, and varying the density dependence
of the corresponding symmetry energy within the limits determined by isovector
properties of finite nuclei, we estimate the constraints on the
core-crust transition density and pressure of neutron stars:
$0.086 \ {\rm fm}^{-3} \leq n_t < 0.090 \ {\rm fm}^{-3}$ and
$0.3\ {\rm MeV \ fm}^{-3} < P_t \leq 0.76 \ {\rm MeV \ fm}^{-3}$.

\vspace{0.3cm}

PACS number(s): 21.30.Fe, 21.60.Jz, 26.60.Gj, 26.60.Kp  \\

Keywords: Nuclear density functional, Equation of state, Neutron
star crust.

\end{abstract}

\section{Introduction}

Neutron stars are extraordinary astronomical laboratories for the
physics of dense neutron-rich nuclear matter
\cite{Shapiro-83,Haensel-07}. They consists of several distinct
layers: the atmosphere, the surface, the crust and the core. The
latter, divided into the outer core and inner core, has a radius
of approximately 10 km and contains most of the star's mass. The
crust, of $\approx 1$ km thickness and containing only a few
percent of the total mass, can also be divided into the outer
crust and inner crust.  Although less exotic and smaller in size
than the core, the crust is nevertheless crucial for the
understanding of the physics of neutron stars. It represents the
interface between the observable surface phenomena and the
invisible core. The structure of the crust can be related to some
peculiar phenomena, such as pulsar glitches, thermal relaxation
after matter accretion, quasi periodic oscillations and
anisotropic surface cooling
\cite{Lattimer-00,Lattimer-01,Link-99}. A very important
ingredient in the study of the structure and various properties of
neutron stars is the equation of state (EOS) of neutron-rich
nuclear matter \cite{Lattimer-07}.

One of the most important prediction of a given EOS is the location of
the inner edge of a neutron star crust. The inner crust comprises the
region from the density at which neutrons drip-out of nuclei,
to the inner edge separating the solid crust from the homogeneous
liquid core. At the inner edge, in fact,
a phase transition occurs from the high-density homogeneous
matter to the inhomogeneous matter at lower densities.
In the transitional region nuclear matter exhibits instability
against clusterization into a two-phase system: neutron-rich  nuclei
immersed in dripped neutrons (and sometimes protons). As nuclei
are arranged in a lattice, they form solid state crust covering the
star's core, which is considered to be a homogeneous liquid
\cite{Kubis-07}. The uniform matter is nearly pure neutron matter,
with  a proton fraction of a few percent, determined by the condition of
$\beta$-equilibrium. The transition density takes its critical value
$n_c$ when the uniform neutron-proton-electron matter (npe) becomes
unstable with respect to the separation into two coexisting phases
(one corresponding to nuclei, the other to a nucleonic sea) \cite{Lattimer-07}.

While the density at which neutrons drip-out of nuclei is rather
well determined, the transition density $n_t$ at the inner edge is
much less certain due to our insufficient knowledge of the
EOS of neutron-rich nuclear matter. The value of
$n_t$ determines the structure
of the inner part of the crust. If sufficiently high, it is
possible for non-spherical phases, with rod- or plate-like nuclei,
to occur before the nuclei dissolve. If $n_t$ relatively low, then
matter makes a direct transition from spherical nuclei to
uniform nucleonic fluid. The extent to which non-spherical
phases occur will have important consequences for other
properties that are determined by the solid crust \cite{Pethick-95b}.

In general, the determination of the transition density $n_t$
itself is a very complicated problem because the inner crust may
have a very complicated structure. A well established approach is
to find the density at which the uniform liquid first becomes
unstable against small-amplitude density fluctuations, indicating
the formation of nuclear clusters. This approach includes the
dynamical method
\cite{Pethick-95b,Baym-71a,Baym-71b,Pethick-95a,Douchin-00,
Oyamatsu-07,Ducoin-07, Xu-09-1,Xu-09-2}, the thermodynamical
method \cite{Lattimer-07,Kubis-07,Kubis-04,Worley-08}, and the
random phase approximation (RPA) \cite{Horowitz-01,Carriere-03}.

All theoretical studies have shown that the
core-crust transition density and pressure are very sensitive to the
density dependence of the nuclear matter symmetry energy. The
EOS of neutron-rich nuclear matter has been constrained by using
results from heavy-ion reaction studies \cite{LCK08}. In particular,
it has been shown that the $E_{sym}(\rho)$ constrained in the
same sub-saturation density range
as the neutron star crust by the isospin diffusion data in heavy-ion
collisions at intermediate energies \cite{Tsa04,Che05a,LiBA05c},
limits the transition density
and pressure to $0.040$ fm$^{-3}$ $\leq \rho _{t}\leq 0.065$
fm$^{-3}$ and $0.01$ MeV/fm$^{3}$ $\leq P_{t}\leq 0.26$
MeV/fm$^{3}$, respectively . These constrained values
appear to be significantly lower than their
fiducial values currently used in the literature.
In a very recent study \cite{Xu-09-2}, the core-crust transition density
and pressure have been systematically analyzed using the dynamical
and thermodynamical methods with a modified
Gogny (MDI) and a set of $51$ different Skyrme interactions. Most of
these interactions predict values for the transition density
and pressure that are considerably higher than the intervals cited above.

In a recent work \cite{Niksic-08} we have explored a particular class of
empirical relativistic nuclear energy density functionals, with parameters adjusted
to experimental binding energies of a large set of axially deformed nuclei.
Starting from microscopic nucleon self-energies in nuclear matter, and empirical
global properties of the nuclear matter equation of state, the coupling
parameters of the functional have been determined in a careful comparison of the
predicted binding energies with data, for a set of 64 axially deformed nuclei in the
mass regions $A\approx 150-180$ and $A\approx 230-250$. The isovector channel,
in particular, has been carefully adjusted to reproduce available data in
medium-heavy and heavy nuclei, including neutron-skin thickness
and excitation energies of isovector dipole resonances. It will be interesting, therefore,
to apply this class of relativistic density functionals in a systematic investigation of the
transition density $n _{t}$ and pressure $P_{t}$ at the inner
edge separating the liquid core from the solid crust of neutron stars.
In the present study the thermodynamical method will be used.

In recent years, there has been an increased interest in studies
of the relationship between the size of the neutron-skin in heavy
nuclei, and the symmetry energy at subsaturation densities
\cite{Horowitz-01,Brown-00,Furnstahl-02,Dieperink-03,Steiner-05,
Todd-05,Chen-05,Sammarruca-09a,
Vidana-09,Centelles-09,Yoshida-04,Klimkiewicz-07}. Studies have
also been reported on the correlation between the size of the
neutron-skin and properties of a neutron star crust. This was
pioneered by Horowitz et. al. \cite{Horowitz-01}, who used the
Random Phase Approximation  based on the Relativistic Mean-Field
(RMF) framework for nuclear matter and finite nuclei. An almost
linear correlation was established between the predicted
core-crust transition density $n_t$ and the size of the
neutron-skin. More recently such studies have been carried out by
by Xu et. al. \cite{Xu-09-1,Xu-09-2}, confirming this linear
correlation.

The article is organized as follows. In Sec. II we review the thermodynamical
method used for locating the inner edge of a neutron star crust. Sec. III
contains a brief description of relativistic density functionals that will be
used to analyze the constraints on the core-crust transition density and
pressure of neutron stars. The results are presented and discussed in
Sec. IV, and Sec. V summarizes the present study.

\section{The Thermodynamical Method}

The core-crust interface corresponds to the phase transition
between nuclei and uniform nuclear matter. The uniform matter is
nearly pure neutron matter, with a proton fraction of just a few
percent determined by the condition of beta equilibrium. Weak
interactions conserve both baryon number and charge
\cite{Lattimer-07}, and from the first low of thermodynamics, at
temperature $T=0$ (for details see the Appendix):
\begin{equation}
{\rm d}u=-P{\rm d}v-\hat{\mu}{\rm d}q, \label{u-1}
\end{equation}
where $u$ is the internal energy per baryon, $v=1/n$, $q=x-Y_e$ is
the charge density, $x$ is the proton fraction in baryonic matter,
$Y_e$ is the electron density, and in $\beta$-equilibrium
$\hat{\mu}=\mu_n-\mu_p=\mu_e$. The stability of the uniform phase
requires that $u(v,q)$ is a convex function \cite{Callen-85}. This
condition leads to the following two constraints for the pressure
and the chemical potential
\begin{equation}
-\left(\frac{\partial P}{\partial v}\right)_q-\left(\frac{\partial
P}{\partial q}\right)_v \left(\frac{\partial q}{\partial
v}\right)_{\hat{\mu}}>0, \label{cond-1}
\end{equation}
\begin{equation}
-\left(\frac{\partial \mu}{\partial q}\right)_v>0. \label{cond-2}
\end{equation}
It is assumed that the total internal energy per baryon $u(v,q)$
can be decomposed into baryon ($E_N$) and electron ($E_e$)
contributions
\begin{equation}
u(v,q)=E_N(v,q)+E_e(v,q). \label{u-2}
\end{equation}
In this work the well know parabolic approximation is used for
the baryon energy $E_N(v,q)$
\begin{equation}
E_N(v,q)\simeq V(v)+E_{sym}(v)(1-2x)^2\; ,
\label{EN-1}
\end{equation}
where higher-order terms in the isospin asymmetry $\delta=1-2x$ are
neglected. In Eq.~({\ref{EN-1}) $V(v)$ denotes the energy of symmetric
nuclear matter, and $E_{sym}(v)$ is given  by
\begin{equation}
E_{sym}(v)\simeq E_N(v,q(x=0))-E_N(v,q(x=0.5)). \label{Esym-1}
\end{equation}
The electron contribution to the total energy reads
\begin{equation}
E_e =\frac{3}{4}Y_e \mu_e. \label{Ee-1}
\end{equation}
The condition of $\beta$-equilibrium leads to the relation
\cite{Prakash-94,Moustakidis-07}
\begin{equation}
\hat{\mu}=-\left(\frac{\partial E_N}{\partial
x}\right)_n=4E_{sym}(n)(1-2x)=\hbar c(3\pi^2 n_e)^{1/3}.
\label{mu-1a}
\end{equation}
From the relation
\begin{equation}
q=x-Y_e=x-n_e/n, \qquad n_e=\frac{\mu_e^3}{\hbar^3c^3 3
\pi^2}=\frac{\hat{\mu}^3}{\hbar^3c^3 3 \pi^2}, \label{q-1}
\end{equation}
and Eq.~(\ref{mu-1}), it follows that
\begin{equation}
-\left(\frac{\partial q}{\partial \hat{\mu}}
\right)_{v(n)}=-\left(\frac{\partial x}{\partial
\hat{\mu}}\right)_{v(n)}+\frac{1}{n}\left(\frac{\partial
n_e}{\partial
\hat{\mu}}\right)_{v(n)}=\frac{1}{8E_{sym}(n)}+\frac{\hat{\mu}^2}{n\hbar^3c^3
\pi^2}=\frac{1}{8 E_{sym}(n)}+\frac{3Y_e}{\hat{\mu}}. \label{AA-1}
\end{equation}
The inequality (\ref{cond-1}) is equivalent to
\begin{equation}
-\left(\frac{\partial P}{\partial v}\right)_{\hat{\mu}}>0.
\label{inq-1}
\end{equation}
The electron contribution to the pressure  $P_e$ is a function of
the chemical potential $\hat{\mu}=\mu_e$ only
\begin{equation}
P_e=\frac{1}{12\pi^2}\frac{\mu_e^4}{(\hbar c)^3}\; , \label{Pe-1}
\end{equation}
and thus the inequality (\ref{inq-1}) can be written as
\begin{equation}
-\left(\frac{\partial P_b}{\partial v}\right)_{\hat{\mu}}>0 ,
\qquad {\rm or}\qquad  n^2 \left(\frac{\partial P_b}{\partial
n}\right)_{\hat{\mu}}>0. \label{inq-2}
\end{equation}
In general the baryonic pressure $P_b$ is a function of both $n$
and $x$, but for a fixed $\hat{\mu}$ (see also Eq.~(\ref{mu-1})
$x=x(n)$, so that $P=P(n,x(n))$, and therefore
\begin{equation}
 n^2 \left(\frac{\partial P_b}{\partial
n}\right)_{\hat{\mu}}=n^2 \left[\frac{{\rm d }P_b}{{\rm
d}n}+\frac{\partial P_b}{\partial x}\left(\frac{\partial
x}{\partial n} \right)_{\hat{\mu}} \right]. \label{Pb-2}
\end{equation}
Now, considering that $\hat{\mu}=\hat{\mu}(n,x)$, it follows that
\begin{equation}
d\hat{\mu}= \left(\frac{\partial \hat{\mu}}{\partial n}\right)_x
{\rm d}n+ \left(\frac{\partial \hat{\mu}}{\partial x} \right)_n
{\rm d}x \Rightarrow \left(\frac{\partial x}{\partial
n}\right)_{\hat{\mu}}=-\left(\frac{\partial \hat{\mu}}{\partial
n}\right)_x \left(\frac{\partial \hat{\mu}}{\partial x}
\right)_n^{-1}, \label{mu-d-1}
\end{equation}
and making use of Eq.~(\ref{mu-1}), one obtains
\begin{equation}
\left(\frac{\partial x}{\partial n}\right)_{\hat{\mu}}=-
\left(\frac{\partial^2 E_N}{\partial n \partial x}\right)
\left(\frac{\partial^2 E_N}{\partial x^2} \right)^{-1}.
\label{mu-d-2}
\end{equation}
Eq.~(\ref{Pb-2}) now reads
\begin{equation}
 n^2 \left(\frac{\partial P_b}{\partial
n}\right)_{\mu}=n^4 \left[\frac{2}{n}\frac{{\rm d }E_N}{{\rm
d}n}+\frac{{\rm d }^2E_N}{{\rm
d}n^2}-\left(\frac{\partial^2E_N}{\partial n \partial
x}\right)^2\left(\frac{\partial^2 E_N}{\partial x^2} \right)^{-1}
\right]. \label{Pb-3}
\end{equation}

The condition of charge neutrality $q=0$ requires that $x=Y_e$. This is
the case we will consider in the present study. After some algebra, it
can be shown that the conditions of Eqs.~(\ref{Pb-3}) and
(\ref{AA-1}) are equivalent to
\begin{equation}
C_I(n)=n^2\frac{{\rm d}^2V}{{\rm d}n^2}+2n\frac{{\rm
d}V}{dn}+(1-2x)^2\left[ n^2\frac{{\rm d}^2 E_{sym}}{{\rm
d}n^2}+2n\frac{{\rm d}E_{sym}}{{\rm d}n}-2\frac{1}{E_{sym}}\left(n
\frac{{\rm d}E_{sym}}{{\rm d}n} \right)^2 \right]>0, \label{K-I}
\end{equation}
\begin{equation}
C_{II}(n)=-\left(\frac{\partial q}{\partial
\hat{\mu}}\right)_v=\frac{1}{8E_{sym}}+\frac{3Y_e}{\hat{\mu}}>0.
\label{K-II}
\end{equation}
The second inequality (\ref{K-II}) is usually valid. The
transition density $n_t$ is determined from the first inequality
(\ref{K-I}). For a given EOS, the quantity $C_{I}(n)$ is plotted
as a function of the baryonic density $n$, and the equation
$C_I(n_t)=0$ defines the transition density $n_t$.

\subsection*{The quantities $L$ and $K_{sym}$}

The density dependence of the nuclear matter symmetry energy
can be characterized in terms of a few bulk parameters by expanding
it in Taylor series around the saturation density $n_0$
\begin{equation}
E_{sym}(n)=E_{sym}(n_0)+L\left(\frac{n-n_0}{3n_0}\right)+
\frac{K_{sym}}{2}\left( \frac{n-n_0}{3n_0} \right)^2+ {\cal O}(3)
\dots,\label{Esym-expa}
\end{equation}
where $E_{sym}(n_0) \equiv a_4$ is the value of the symmetry energy at
saturation, $L$ is the slope parameter
\begin{equation}
L=3n_0 \left(\frac{\partial E_{sym}(n)}{\partial
n}\right)_{n=n_0}, \label{L-1}
\end{equation}
and the curvature parameter$K_{sym}$  is the isovector correction
to the compression modulus
\begin{equation}
K_{sym}=9n_0^2\left(\frac{\partial^2 E_{sym}(n)}{\partial
n^2}\right)_{n=n_0}. \label{Ksym-1}
\end{equation}
The slope parameter $L$  is related to
$P_0$, the pressure from the symmetry energy for pure neutron
matter at saturation density \cite{Tsang-09}. The symmetry
pressure $P_0$ provides the dominant baryonic contribution to
the pressure in neutron stars at saturation density. It will be
interesting to study the relation between the transition density
$n_t$, and $L$ and $K_{sym}$, as well as to examine the correlations
between $L$ and $K_{sym}$ and the neutron-skin thickness.

The neutron-skin thickness $S$ of a nucleus is defined as the
difference between the root-mean-square radii  of neutron
$\sqrt{\langle r_n^2 \rangle }$ and proton $\sqrt{\langle r_p^2
\rangle }$ distributions
\begin{equation}
S=\sqrt{\langle r_n^2 \rangle }- \sqrt{\langle r_p^2
\rangle}=R_n-R_p. \label{S-1}
\end{equation}
$S$ is sensitive to the density dependence
of the nuclear symmetry energy, particularly the slope parameter $L$
 \cite{Brown-00,Chen-05,Sammarruca-09a}.
More specifically, the slope parameter $L$ has been found to
correlate linearly with the neutron-skin thickness of heavy nuclei
 \cite{Horowitz-01,Todd-05,Chen-05,Centelles-09}.

\subsection*{The pressure at the inner edge of a neutron star crust}

The pressure at the inner edge is an important quantity directly related
to the crustal fraction of the moment of inertia, which
can be measured indirectly from observations of pulsars glitches
\cite{Lattimer-07}. The total pressure is decomposed into baryon
and lepton contributions
\begin{equation}
P(n,x)=P_b(n,x)+P_e(n,x), \label{P-all-1}
\end{equation}
where
\begin{equation}
P_b(n,x)=n^2\frac{{\rm d} E_N}{{\rm d} n}, \qquad
E_N(n,x)=V(n)+(1-2x)^2E_{sym}(n). \label{Pb-1}
\end{equation}
The baryon pressure $P_b$, therefore, is given by
\begin{equation}
P_b(n,x)=n^2\left[\frac{{\rm d} V(n)}{{\rm d} n}+\frac{{\rm d}
E_{sym}}{{\rm d} n}(1-2x)^2 \right]. \label{Pb-2}
\end{equation}
The electrons are considered as a non-interacting Fermi gas.
Their contribution to the total pressure reads
\begin{equation}
P_e(n,x)=\frac{1}{12\pi^2}\frac{\mu_e^2}{(\hbar c)^3}=\frac{\hbar
c}{12 \pi^2}\left(3\pi^2 xn\right)^{4/3}. \label{Pe-2}
\end{equation}
For a given symmetry energy $E_{sym}(n)$,
equation (\ref{mu-1a}) determines the equilibrium proton fraction
$x(n)$. A simple algebra leads to
\begin{equation}
x(n)=\frac{1}{2}-\frac{1}{4}\left([2\beta
(\gamma+1)]^{1/3}-[2\beta (\gamma-1)]^{1/3}\right) , \label{x-1}
\end{equation}
where \[ \beta(n)=3\pi^2 n(\hbar c/4E_{sym}(n))^3 ,\qquad \qquad
\gamma(n) =\left(1+\frac{2 \beta}{27} \right)^{1/2}. \]
%

\section{Relativistic Energy Density Functionals}

The framework of nuclear energy density functionals (NEDF) provides, at present,
the most complete microscopic approach to the rich variety of structure phenomena
in medium-heavy and heavy complex nuclei, including regions of the nuclide chart
far from the valley of $\beta$-stability. By employing global
functionals parameterized by a set of $\approx 10$  coupling constants,  the current
generation of EDF-based models has achieved a high level of accuracy in the description
of ground states and properties of excited states, exotic unstable nuclei, and even
nuclear systems at the nucleon drip-lines.

There are important advantages in using relativistic density functionals, i.e.
functionals with manifest covariance. The most obvious is the natural inclusion of the
nucleon spin degree of freedom, and the resulting nuclear spin-orbit potential
which emerges automatically with the empirical strength in a covariant formulation.
The consistent treatment of large isoscalar, Lorentz scalar and vector self-energies
provides a unique parametrization of time-odd components of the nuclear
mean-field, i.e. nucleon currents. Recently \cite{Niksic-08} we have explored
a particular class of relativistic nuclear energy density functionals in which only
nucleon degrees of freedom are explicitly used in the construction of effective interaction terms.
Short-distance correlations, as well as intermediate and long-range dynamics, are effectively
taken into account in the nucleon-density dependence of the strength functionals of second-order
contact interactions in an effective Lagrangian.

The basic building blocks are the densities and currents
bilinear in the Dirac spinor field $\psi$ of the nucleon:
$\bar{\psi}\mathcal{O}_\tau \Gamma \psi$, with $\mathcal{O}_\tau \in \{1,\tau_i\}$ and
 $\Gamma \in \{1,\gamma_\mu,\gamma_5,\gamma_5\gamma_\mu,\sigma_{\mu\nu}\}$.
Here $\tau_i$ are the isospin Pauli matrices and $\Gamma$ generically denotes the Dirac
matrices. The nuclear ground-state density and energy are determined by the
self-consistent solution of relativistic linear single-nucleon Kohn-Sham equations.
To derive those equations it is useful to construct an
interaction Lagrangian with four-fermion
(contact) interaction terms in the various isospace-space channels:
isoscalar-scalar   $(\bar\psi\psi)^2$,
isoscalar-vector $(\bar\psi\gamma_\mu\psi)(\bar\psi\gamma^\mu\psi)$,
isovector-scalar $(\bar\psi\vec\tau\psi)\cdot(\bar\psi\vec\tau\psi)$,
isovector-vector $(\bar\psi\vec\tau\gamma_\mu\psi)
                         \cdot(\bar\psi\vec\tau\gamma^\mu\psi)$ .
 A general Lagrangian can be written as a power series in the currents
 $\bar{\psi}\mathcal{O}_\tau\Gamma\psi$ and their derivatives, with
higher-order terms representing in-medium many-body correlations.
The Lagrangian considered in Ref.~\cite{Niksic-08}
includes second-order interaction terms, with many-body correlations
(short-distance correlations, as well as
intermediate and long-range dynamics), encoded in
density-dependent coupling functions:
\begin{eqnarray}
\label{Lagrangian}
\mathcal{L} &=& \bar{\psi} (i\gamma \cdot \partial -m)\psi \nonumber \\
     &-& \frac{1}{2}\alpha_S(\hat{n})(\bar{\psi}\psi)(\bar{\psi}\psi)
       - \frac{1}{2}\alpha_V(\hat{n})(\bar{\psi}\gamma^\mu\psi)(\bar{\psi}\gamma_\mu\psi)
     - \frac{1}{2}\alpha_{TV}(\hat{n})(\bar{\psi}\vec{\tau}\gamma^\mu\psi)
                                                                 (\bar{\psi}\vec{\tau}\gamma_\mu\psi) \nonumber \\
    &-&\frac{1}{2} \delta_S (\partial_\nu \bar{\psi}\psi)  (\partial^\nu \bar{\psi}\psi)
         -e\bar{\psi}\gamma \cdot A \frac{(1-\tau_3)}{2}\psi\;.
\end{eqnarray}
In addition to the free-nucleon Lagrangian and the point-coupling interaction terms,
when applied to nuclei, the model must include the coupling of the protons to the
electromagnetic field.
The derivative term in Eq.~(\ref{Lagrangian}) accounts for leading effects of finite-range
interactions that are crucial for a quantitative description of nuclear density distribution,
e.g. nuclear radii.  Eq.~(\ref{Lagrangian}) includes only one isovector term, i.e. the
isovector-vector interaction because, although the isovector strength has a relatively
well-defined value, the distribution between the scalar and vector channels is not
determined by ground-state data.

The strength and density dependence of the interaction terms of
the Lagrangian Eq.~(\ref{Lagrangian}) are parameterized as follows:
\begin{eqnarray}
\alpha_S(n)&=& a_S + (b_S + c_S x)e^{-d_S x},\nonumber\\
\alpha_V(n)&=& a_V +  b_V e^{-d_V x},\\
\alpha_{TV}(n)&=& b_{TV} e^{-d_{TV} x}\nonumber
\label{parameters}
\end{eqnarray}
where $x=n/n_{0}$, and $n_{0}$ denotes the nucleon density
at saturation in symmetric nuclear matter. The set of 10 parameters has been
adjusted in a multistep parameter fit exclusively to the experimental
masses of 64 axially deformed nuclei in the mass regions $A\approx
150-180$ and $A\approx 230-250$. The resulting best-fit functional DD-PC1
has been further tested in calculations of binding energies, charge
radii, deformation parameters, neutron-skin thickness, and excitation
energies of giant monopole and dipole resonances. The nuclear matter
equation of state, corresponding to DD-PC1, is characterized by the
following properties at the saturation point: nucleon density
$\rho_{sat}=0.152~\textnormal{fm}^{-3}$, volume energy $a_v=-16.06$
MeV, symmetry energy $a_4 = 33$ MeV,
and the nuclear matter compression modulus $K_{nm} = 230$ MeV.

\section{Results and Discussions}

To adjust the functional DD-PC1, in  Ref.~\cite{Niksic-08} sets of effective
interactions with different values of the volume $a_v$, surface $a_s$, and
symmetry energy $a_4$ in nuclear matter were generated, and the corresponding binding
energies of deformed nuclei with $A\approx 150-180$ and $A\approx 230-250$ were
analyzed. The nuclear matter saturation density, the Dirac mass,
and the compression modulus, were kept fixed: $n_0 =0.152~\textnormal{fm}^{-3}$
in accordance with values predicted by most modern relativistic mean-field models,
$m^*_D = 0.58 m$ in the narrow interval of values allowed by the empirical
energy spacings between spin-orbit partner states in finite nuclei, and
$K_{nm} = 230$ MeV to reproduce experimental excitation energies of isoscalar giant
monopole resonances in relativistic (Q)RPA calculations.

Nuclear structure data do not constrain the
nuclear matter EOS at high nucleon densities. Therefore, two additional points on
the $E(\rho)$ curve in symmetric matter were fixed to the microscopic EoS of Akmal,
Pandharipande and Ravenhall \cite{Akmal-98}, based on the Argonne V$_{18}$
NN potential and the UIX three-nucleon interaction. This EOS has extensively been
used in studies of high-density nucleon matter and neutron stars. At almost
four times nuclear matter saturation density, the point $n =0.56~\textnormal{fm}^{-3}$
with $E/A = 34.39$ MeV was chosen and,
to have an overall consistency, one point at low density:
$n =0.04~\textnormal{fm}^{-3}$ with $E/A = -6.48$ MeV
(cf. Table VI of Ref.~\cite{Akmal-98}).

The calculated binding energies of finite nuclei are very
sensitive to the choice of the nuclear matter volume energy
coefficient $a_v$. In fact, one of the important results of
analysis of deviations between calculated and experimental masses
(mass residuals) of Ref.~\cite{Niksic-08}, is the pronounced
isospin and mass dependence of the residuals on the nuclear matter
volume energy at saturation. To reduce the absolute mass residuals
to less than 1 MeV, and to contain their mass and isotopic
dependence, $a_v$ had to be constrained to a narrow interval of
values: $-16.04$ MeV $\leq a_v \leq  -16.08$ MeV. Experimental
masses do not place very strict constraints on the parameters of
the expansion of $E_{sym}(n)$ (cf. Eq.~(\ref{Esym-expa})), but
self-consistent mean-field calculations show that binding energies
can restrict the values of $E_{sym}$ at nucleon densities somewhat
below saturation density, i.e. at $ n  \approx
0.1~\textnormal{fm}^{-3}$. Additional information on the symmetry
energy can be obtained from data on neutron-skin thickness and
excitation energies of giant dipole resonances. Recent studies
have shown that relativistic effective interactions with volume
asymmetry $a_4$ in the range $31\;\textnormal{MeV} \le a_4 \le
35\; \textnormal{MeV}$ predict values for neutron-skin thickness
that are consistent with data, and reproduce experimental
excitation energies of isovector giant dipole
resonance~\cite{VNR.03}. Therefore, in the construction of the
functional DD-PC1 in Ref.~\cite{Niksic-08}, the volume asymmetry
was held fixed at $a_4=33$ MeV, and the symmetry energy at a
density that corresponds to an average nucleon density in finite
nuclei: $\langle n \rangle =0.12\;\textnormal{fm}^{-3}$ was
varied. The quantity $E_{sym}(n=0.12\;\textnormal{fm}^{-3})$ will
be denoted $\langle S_2 \rangle$.

Starting from the relativistic energy density functional DD-PC1, in this work we
examine the sensitivity of the core-crust transition density $n_t$ and pressure
$P_t$ of neutron stars, on the density dependence of the corresponding
symmetry energy of nucleonic matter. In Ref.~\cite{Niksic-08} the value of
$\langle S_2 \rangle$ was varied in a rather narrow interval of values
27.6 MeV $\le \langle S_2 \rangle \le $ 28.6 MeV, constrained by
the empirical values of binding energies and ground-state isovector properties
of finite nuclei. Fig.~\ref{Fig_A} displays the corresponding symmetry energy
curves $E_{sym}$ as a function of the baryon density $n$. For $a_v=-16.06$
meV (DD-PC1) the minimum $\chi^2$ deviation of the theoretical binding
energies from data is obtained when $\langle S_2 \rangle = 27.8$ MeV.

Table \ref{t:1} and Fig.~\ref{Fig_B} display the values of the transition density
$n_t$ (in fm$^{-3}$) and transition pressure $P_t$ (in Mev$\cdot$fm$^{-3}$),
calculated in the thermodynamical model, as functions of $<S_2>$ for three values
of the nuclear matter volume energy coefficient $a_v$.
For a given value of the parameter $a_v$, the values of $n_t$ rise with increasing
 $<S_2>$, whereas the opposite is found for the values of $P_t$. For the considered
 interval of $<S_2>$, however, the changes are small. An increase of
$3.5 \%$ in $<S_2>$ leads to an increase of
$1.5 \%$ in the value of $n_t$. The transition pressure exhibits a somewhat more
pronounced  dependence (the corresponding decrease is around $16-20 \%$).
Both $n_t$ and $P_t$ display a negligible dependence on $a_v$, even though
$a_v = -16.02$ MeV and $a_v = -16.14$ MeV lie outside the interval of values
for which the absolute deviations between calculated and experimental masses are
smaller than 1 MeV.

In Fig.~\ref{Fig_C} we plot the transition pressure $P_t$ as a
function of the transition density $n_t$ for the three sets of
nuclear matter EOS and symmetry energy described above, in
comparison with results of recent calculations performed using
an isospin and momentum-dependent modified
Gogny effective interaction (MDI)
\cite{Xu-09-2,Krastev-010}. The different values of
the parameter  $x$ in the MDI model correspond to various
choices of the density dependence of the nuclear
symmetry energy. In Refs.~\cite{LiBA05c,LiBA05d} it has
been shown that only $-1 \leq x \leq 0$ leads to a density dependence
of the symmetry energy in the sub-saturation density region that is
consistent with isospin diffusion data and the empirical value of
the neutron-skin thickness in $^{208}$Pb. In addition to the
MDI EOS, in Fig.~\ref{Fig_C} we also show the result
obtained by Akmal et al. \cite{Akmal-98} with the $A18+\delta v+UIX^*$ interaction (ARP),
and the value obtained in the recent Dirac-Brueckner-Hartree-Fock (DBHF) calculation
\cite{Sammarruca-09a} with the Bonn B One-Boson-Exchange (OBE)
potential (DBHF+Bonn B) \cite{Machleidt-89}.

A distinctive feature of the present analysis is the
narrow interval of allowed values $(n_t,P_t)$ that results
from the rather stringent constraints on the parameter
$<S_2>$. The effect of varying the volume energy at saturation
 $a_v$ is almost negligible. The present results for $n_t$ and $P_t$
 lie in  the region constrained by the measure of the current uncertainty
in the density dependence of the symmetry energy
\cite{Lattimer-07}, and are found very close to the result of
Akmal et al. \cite{Akmal-98}. We note that all the results shown
in Fig.~\ref{Fig_C} are obtained using the parabolic approximation for
the EOS of isospin-asymmetric nuclear matter.
The transition density and pressure have also been estimated using the full
equation of state and employing both the dynamical and thermodynamical
methods \cite{Xu-09-1,Xu-09-2}.

As explained above, the rather narrow interval of  $<S_2>$, for this type of
nuclear energy density functionals, has been constrained by
the empirical values of binding energies and ground-state isovector properties
of finite nuclei. The symmetry energy at saturation density, $a_4=33$ MeV,
was fixed in Ref.~\cite{Niksic-08} to obtain the best results for the neutron-skin
thickness in Sn isotopes and $^{208}$Pb, and for the excitation energies of
isovector dipole resonances. However, because of large experimental uncertainties,
especially for the neutron-skin thickness, good agreement with data can also be
obtained for other values of $a_4$. This is shown in Fig.~\ref{Fig_D}, where we
plot the predictions for the differences between neutron and proton $rms$ radii
of Sn and Pb isotopes, in comparison with available data \cite{Kra.99,SH.94,Kra.94},
for different choices of the symmetry energy at saturation density. The self-consistent
mean-field calculations have been performed using the relativistic Hartree-Bogoliubov
(RHB) model \cite{VALR.05}, with pairing correlations described by the pairing part
of the finite-range Gogny interaction. The isoscalar channel of the particle-hole
interaction corresponds to the DD-PC1 functional, and in the isovector channel
 $<S_2>$ is kept fixed at 27.8 MeV (DD-PC1), whereas  $a_4$ is varied in the
 interval between 30 MeV and 35 MeV. The corresponding symmetry energy as
 a function of the nucleon density is shown in Fig.~\ref{Fig_E}.

 We notice, therefore, that by keeping $<S_2>$ constant and varying $a_4$ in the
 interval between 30 MeV and 35 MeV,  the density dependence of the symmetry energy
 can be modified in a controlled way, i.e. the corresponding energy density functionals
 still reproduce ground-state properties of finite nuclei in fair agreement with data. In
 Table \ref{t:2} and Fig.~\ref{Fig_F} we display the corresponding values of the transition density
$n_t$ (in fm$^{-3}$) and transition pressure $P_t$ (in
Mev$\cdot$fm$^{-3}$) as functions of $a_4$ for three values of the
nuclear matter volume energy coefficient $a_v$. The transition
pressure $P_t$ as a function of the transition density $n_t$ for
the three sets of nuclear matter EOS and symmetry energy is
plotted in Fig.~\ref{Fig_G}. Not surprising, considering the
symmetry energy curves of Fig.~\ref{Fig_E}, the constraints on
$n_t$  and $P_t$ have been relaxed in this case, and the allowed
values span a much larger interval of values compared to the
restricted variation of $<S_2>$ shown in Figs.~\ref{Fig_A} and
\ref{Fig_B}.

To be able to compare the present results for the transition density
and transition pressure with recent studies \cite{Xu-09-2}, in Fig.~\ref{Fig_H}
we plot the calculated values of $n_t$ and $P_t$ as functions of the slope
parameter of the symmetry energy (cf. Eq.~(\ref{L-1})), for the two sets of
effective interactions described above. $n_t$ is a monotonously decreasing,
and $P_t$ monotonously increasing function of $L$. In the small interval
of $L$ values determined by the variation of $<S_2>$ between 27.6 MeV and
28.6 MeV, both $n_t$ and $P_t$ display a linear dependence on $L$. In the
much larger interval determined by the variation of $a_4$ from 30 MeV to
35 MeV, a weak parabolic dependence of $n_t$ and $P_t$ is found.
We note that transport model studies of the isospin diffusion data
in heavy-ion reactions have constrained the slope parameter L
to the values $88 \pm 25$ MeV \cite{Xu-09-2}. Considering that
we can also, most probably, exclude the value $a_4 = 30$ MeV
for the asymmetry at saturation density
(cf. Figs.~\ref{Fig_D} and \ref{Fig_F}), because it implies an unrealistically
small value of $< 0.1$ fm for the neutron-skin thickness of  $^{208}$Pb,
the present analysis places the following constraints on the
core-crust transition density and pressure of neutron stars:
$0.086 \ {\rm fm}^{-3} \leq n_t < 0.090 \ {\rm fm}^{-3}$ and
$0.3\ {\rm MeV \ fm}^{-3} < P_t \leq 0.76 \ {\rm MeV \ fm}^{-3}$.

Finally, in Fig.~\ref{Fig_I} we compare the present prediction for the range of
values of the transition density $n_t$ with the results of Horowitz and Piekarewicz who,
in Ref.~\cite{Horowitz-01}, also used the framework of relativistic mean-field
effective interactions to study the relationship between the neutron-skin thickness of
a heavy nucleus and the properties of neutron star crusts. Starting from the NL3
meson-exchange effective interaction \cite{NL3}, the density dependence of the
symmetry energy was varied by adding nonlinear couplings between the isoscalar
and isovector mesons to the original interaction. The variation was carried out in
such a way to enhance the changes in the neutron density and neutron-skin thickness,
while keeping small the corresponding changes in the binding energy and proton
density distribution. For the solid crust of a neutron star, the effective RMF interactions
were used in a simple RPA calculation of the transition density below which uniform
neutron-rich matter becomes unstable against small amplitude density fluctuations.
The resulting transition densities are plotted in  Fig.~\ref{Fig_I} as a function of the predicted
difference between neutron and proton $rms$ radii in $^{208}$Pb. This inverse
correlation was parameterized \cite{Horowitz-01}
\begin{equation}
n_t\approx 0.16-0.39(R_n-R_p), \label{Hor-1}
\end{equation}
with the skin thickness expressed in fm. In the present analysis,
using a different type of relativistic effective interactions and
varying the density dependence of the symmetry energy by
explicitly modifying $<S_2>$ or $a_4$, we find a much weaker
dependence $n_t$ on the neutron-skin thickness of $^{208}$Pb.


\section{Summary}
The framework of  relativistic nuclear energy functionals has been
employed to analyze and constrain the transition density $n_t$ and
pressure $P_t$ at the inner edge between
the liquid core and the solid crust of a neutron star, using the
thermodynamical method. Starting from a class of energy density
functionals carefully adjusted to experimental masses of finite
nuclei, we have examined the sensitivity of the core-crust transition
density $n_t$ and pressure $P_t$ on the density dependence of
corresponding symmetry energy of nucleonic matter. The limits of
variation of the density dependence of the symmetry energy are
determined by isovector properties of finite nuclei: the thickness
of the neutron-skin and the excitation energies of isovector giant
dipole resonances. Instead of an unrestricted variation of the
parameters of the Taylor expansion of the symmetry energy
around the saturation density of nuclear matter, that is the
slope parameter and the isovector correction to the compression
modulus, we modify the density dependence by varying the
value of the nuclear symmetry energy at a point  somewhat
below the saturation density $\langle S_2 \rangle$
(the symmetry energy at $n=0.12$fm$^{-3}$), and
at the saturation density $a_4$ (the symmetry energy at
$n=0.152$fm$^{-3}$, the saturation density for this class
of relativistic density functionals). In the former case,
for a given value of the volume energy coefficient $a_v$,
$\langle S_2 \rangle$  has been varied in a rather narrow
interval of values 27.6 MeV $\le \langle S_2 \rangle \le $ 28.6 MeV
determined by a fit to the experimental binding energies. We have
found that an increase of $3.5 \%$ in
$<S_2>$ leads to an increase of $1.5 \%$ in the value of $n_t$
while $P_t$ exhibits a somewhat more pronounced dependence (the
corresponding decrease is around $16-20 \%$).
Both $n_t$ and $P_t$ display a negligible dependence on $a_v$.
The variation of the parameter $a_4$ has been in the
range of values: 30 MeV $\le a_4  \le $ 35 MeV, allowed by the empirical
thickness of the neutron-skin and excitation energies of isovector
dipole resonances, for a fixed value of $<S_2>$.
Again, there is virtually no dependence on $a_v$, but now both
$n_t$ and $P_t$ span much wider intervals. We have also examined
the dependence of $n_t$ and $P_t$ on the slope parameter of the
symmetry energy $L$, for the two sets of effective interactions
described above. For the empirical range of the slope parameter
$88 \pm 25$ MeV, and comparing the calculated values of the
neutron-skin thickness with available data for Sn isotopes and
$^{208}$Pb, we have deduced the following constraints on the
core-crust transition density and pressure of neutron stars:
$0.086 \ {\rm fm}^{-3} \leq n_t < 0.090 \ {\rm fm}^{-3}$ and
$0.3\ {\rm MeV \ fm}^{-3} < P_t \leq 0.76 \ {\rm MeV \ fm}^{-3}$.

The present study will be extended to include also the
dynamical and the random phase approximation approaches to
analyze the nuclear constraints on the core-crust transition
density and pressure of neutron stars. Interesting issues for
future study will also be finite temperature and neutrino trapping
effects. Work along these lines is in progress.

\section*{Acknowledgments}
This work was supported in part by MZOS-project 1191005-1010, and
by the DFG cluster of excellence \textquotedblleft Origin and
Structure of the Universe\textquotedblright\
(www.universe-cluster.de). T.N. acknowledges support from the
Croatian National Foundation for Science. D. Vretenar would like
to acknowledge the support from the Alexander von Humboldt
Foundation.

\section*{Appendix}

\subsubsection*{Proof of the equality: \mbox{\boldmath$ {\rm du}=-P{\rm d}v-\hat{\mu}{\rm dq}$}}
From the first low of  thermodynamic, at temperature $T=0$
\begin{equation}
{\rm U}=-P{\rm d V}+\sum_{i}\mu_i{\rm dN_i}, \label{DU-1}
\end{equation}
where $U$ is the total energy of  $N=\sum_iN_i$ particles
in a volume $V$. Here $i=p,n,e$, and the number
of baryons is $N_b=N_n+N_p$. Dividing Eq.~(\ref{DU-1}) by the
baryon number $N_b$, one obtains
\begin{equation}
{\rm u}=-P{\rm d }v+\sum_{i}\mu_i{\rm dY_i}\; , \label{DU-2}
\end{equation}
with
\begin{equation}
\sum_i\mu_idY_i=\mu_n{\rm Y}_n +\mu_p{\rm Y}_p+\mu_e{\rm Y}_e,
 \label{DU-3}
\end{equation}
and
\begin{equation}
Y_n=1-Y_p. \label{DU-4}
\end{equation}
A simple algebra leads to the following relation
\begin{eqnarray}
\sum_i\mu_idY_i&=&-(\mu_n-\mu_p){\rm d Y}_p+\mu_e{\rm d
Y}_e=-\hat{\mu}{\rm d Y}_p+\hat{\mu}{\rm d Y}_e \nonumber \\
& =&-\hat{\mu}({\rm dY}_p-{\rm d Y}_e)=-\hat{\mu}{\rm d q},
\label{DU-5}
\end{eqnarray}
where we have used the equalities
\begin{equation}
\hat{\mu}=\mu_n-\mu_p=\mu_e,\qquad  {\rm and} \qquad q=Y_p-Y_e.
\label{DU-6}
\end{equation}
Eq.~({\ref{DU-2}) therefore takes the form
\begin{equation}
{\rm du}=-P{\rm d}v-\hat{\mu}{\rm dq}. \label{DU-7}
\end{equation}

\subsubsection*{Convexity of the function $u(v,q)$}
Let us consider the function $u(v,q)$  and the
determinant
\begin{equation}
D=\left| \begin{array}{rr}
\frac{\partial ^2 u}{\partial v^2}  & \frac{\partial u^2}{\partial v \partial q}  \\
\frac{\partial u^2}{\partial q \partial v} & \frac{\partial ^2 u}{\partial q^2} \\
\end{array} \right|.
\label{det-1}
\end{equation}
The differential ${\rm d}u(u,q)$ reads
\begin{equation}
{\rm d}u(u,q)=\left(\frac{\partial u}{\partial v}\right)_q {\rm
d}v+ \left(\frac{\partial u}{\partial q} \right)_v {\rm
d}q.\label{du-1}
\end{equation}
From Eqs.~(\ref{DU-7}) and (\ref{du-1}) it follows that
\begin{equation}
P=-\left(\frac{\partial u}{\partial v}\right)_q, \qquad
\hat{\mu}=-\left(\frac{\partial u}{\partial q} \right)_v.
\label{P-mu}
\end{equation}
The determinant Eq.~(\ref{det-1}) takes the form
\begin{equation}
D=\left| \begin{array}{rr}
-\left(\frac{\partial P}{\partial v}\right)_q  & -\left(\frac{\partial P}{\partial q}\right)_v  \\
-\left(\frac{\partial \hat{\mu}}{\partial v}\right)_q & -\left(\frac{\partial \hat{\mu}}{\partial q}\right)_v \\
\end{array} \right|.
\label{det-1}
\end{equation}
The requirement of convexity for the  function $u(v,q)$ leads to the
following two sets of inequalities\\
Case 1:
\begin{equation}
D>0, \qquad -\left(\frac{\partial P}{\partial v}\right)_q >0.
\label{Inq-1}
\end{equation}
Case 2:
\begin{equation}
D>0,\qquad  -\left(\frac{\partial \hat{\mu}}{\partial q}\right)_v
>0. \label{Inq-2}
\end{equation}
The second  set can written in the form
\begin{equation}
\left(\frac{\partial P}{\partial v}\right)_q \left(\frac{\partial
\hat{\mu}}{\partial q}\right)_v - \left(\frac{\partial P}{\partial
q}\right)_v \left(\frac{\partial \hat{\mu}}{\partial
v}\right)_q>0, \label{K-1}
\end{equation}
\begin{equation}
-\left(\frac{\partial \hat{\mu}}{\partial q}\right)_v >0.
\label{K-2}
\end{equation}
Dividing the inequality (\ref{K-1}) by the positive quantity
$-\left(\frac{\partial \hat{\mu}}{\partial q}\right)_v$, and
considering that the differential of the function $\hat{\mu}(v,q)$
is given by
\begin{equation}
d\hat{\mu}= \left(\frac{\partial \hat{\mu}}{\partial v}\right)_q
{\rm d}v+ \left(\frac{\partial \hat{\mu}}{\partial q} \right)_v
{\rm d}q,\label{mu-1}
\end{equation}
and therefore for a fixed $\hat{\mu}$
\begin{equation}
\left(\frac{\partial \hat{\mu}}{\partial q}
\right)_v=-\left(\frac{\partial \hat{\mu}}{\partial v}   \right)_q
\left(\frac{\partial q}{\partial v}   \right)^{-1}_{\hat{\mu}},
\label{mu-2}
\end{equation}
one obtains the following inequality
\begin{equation}
-\left(\frac{\partial P}{\partial v}\right)_q -
\left(\frac{\partial P}{\partial q}\right)_v \left(\frac{\partial
q}{\partial v}\right)_{\hat{\mu}}>0. \label{K-1b}
\end{equation}

\newpage


\newpage
 \begin{table}[h]
\begin{center}
\caption{The values of the transition density $n_t$ (in fm$^{-3}$) and
transition pressure $P_t$ (in Mev$\cdot$fm$^{-3}$),
calculated in the thermodynamical model, as functions of $<S_2>$, for
$a_4 = 33$ MeV and three values
of the nuclear matter volume energy coefficient $a_v$.
}
 \label{t:1}
\vspace{0.5cm}
\begin{tabular}{|cccccccc|}
\hline
      & $a_v=-16.02 \ {\rm MeV}$  &      & $a_v=-16.08 \ {\rm MeV}$  &      &$a_v=-16.14 \ {\rm MeV}$  &  &   \\
   $<S_2> $   & $n_t$ &
$P_t $ & $n_t$ & $P_t $ &
$n_t$ & $P_t$ &    \\
\hline
27.6   & 0.0868  & 0.598  & 0.0867 & 0.590 & 0.0867  & 0.592&  \\
 \hline
 27.8  &  0.0870  & 0.581  & 0.0869  & 0.573 & 0.0869    & 0.576& \\
 \hline
 28.0  &  0.0872   & 0.563  & 0.0871 & 0.556 & 0.0872  & 0.558& \\
 \hline
 28.2 & 0.0875   & 0.544  & 0.0874 & 0.537 & 0.0874  & 0.539 & \\
 \hline
  28.4& 0.0878   & 0.524  & 0.0877 & 0.516 & 0.0877  & 0.519 & \\
 \hline
  28.6& 0.0881   & 0.502  & 0.0880 & 0.495 & 0.0880  & 0.498&  \\
 \hline
\end{tabular}
\end{center}
\end{table}

 \begin{table}[h]
\begin{center}
\caption{The values of the transition density $n_t$ (in fm$^{-3}$) and
transition pressure $P_t$ (in Mev$\cdot$fm$^{-3}$),
calculated in the thermodynamical model, as functions of $a_4$, for
$<S_2> = 27.8$ MeV and three values
of the nuclear matter volume energy coefficient $a_v$.
 }
 \label{t:2}
\vspace{0.5cm}
\begin{tabular}{|cccccccc|}
\hline
      & $a_v=-16.02 \ {\rm MeV}$  &      & $a_v=-16.08 \ {\rm MeV}$  &      &$a_v=-16.14 \ {\rm MeV}$  &  &   \\
   $a_4 ({\rm MeV}) $   & $n_t$ &
$P_t $ & $n_t$ & $P_t $ &
$n_t$ & $P_t$ &    \\
\hline
30   & 0.0922  & 0.091  & 0.0921 & 0.085 & 0.0921  & 0.087&  \\
 \hline
 31  &  0.0899  & 0.303  & 0.0898  & 0.297 & 0.0898    & 0.299& \\
 \hline
 32  &  0.0883   & 0.463  & 0.0882 & 0.456 & 0.0883  & 0.459& \\
 \hline
 33 & 0.0873   & 0.587  & 0.0872 & 0.580 & 0.0873  & 0.582 & \\
 \hline
  34& 0.0866   & 0.682  & 0.0865 & 0.675 & 0.0865  & 0.677 & \\
 \hline
  35& 0.0859   & 0.755  & 0.0859 & 0.748 & 0.0859  & 0.750&  \\
 \hline
\end{tabular}
\end{center}
\end{table}
\newpage
\begin{figure}
\centering
\includegraphics[height=10.0cm,width=10.0cm]{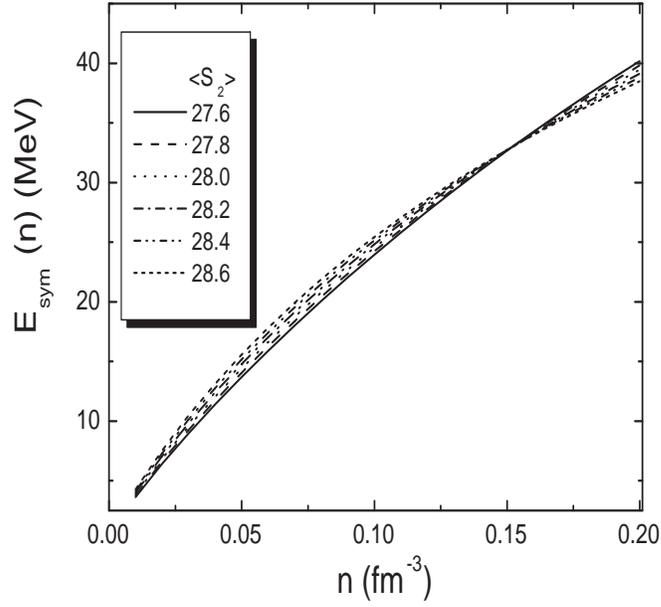}
\caption{The symmetry energy $E_{sym}$ as a function of the
nucleon density $n$ for various values of the parameter $<S_2>$.
The symmetry energy at saturation is $a_4 = 33$ MeV.}
\label{Fig_A}
\end{figure}
\begin{figure}
\centering
\includegraphics[height=8.0cm,width=8.0cm]{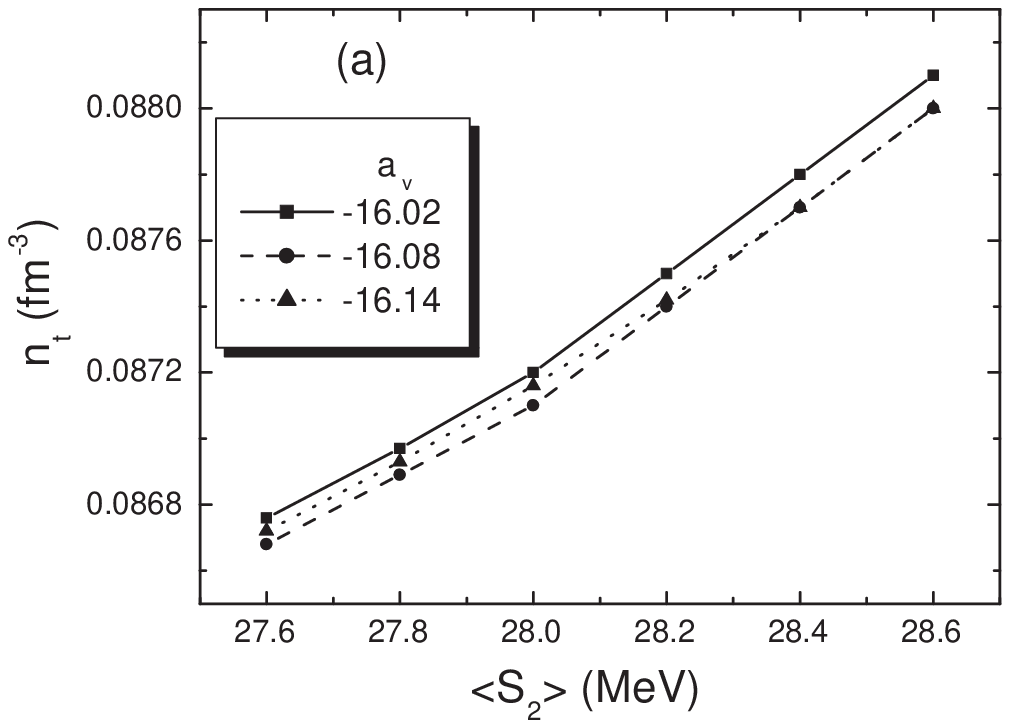}\
\includegraphics[height=8.0cm,width=8.0cm]{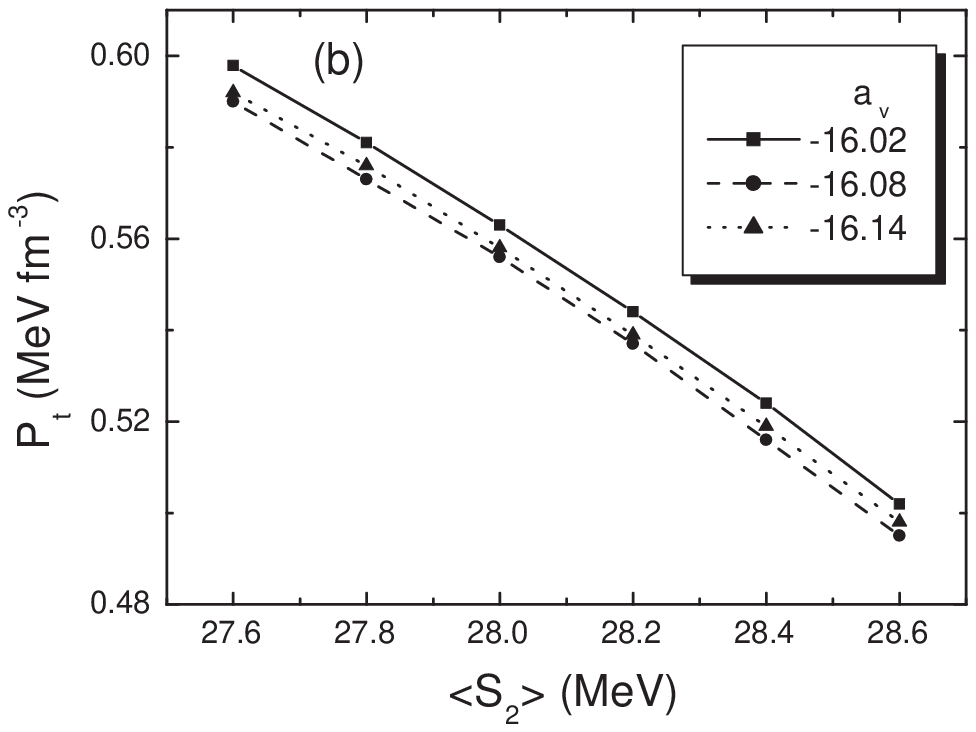}\
\caption{The transition density $n_t$ (a), and the transition
pressure $P_t$ (b), as functions of $<S_2>$ for three values of
the nuclear matter volume energy coefficient $a_v$.}
\label{Fig_B}
\end{figure}
\begin{figure}
\centering
\includegraphics[height=10cm,width=10cm]{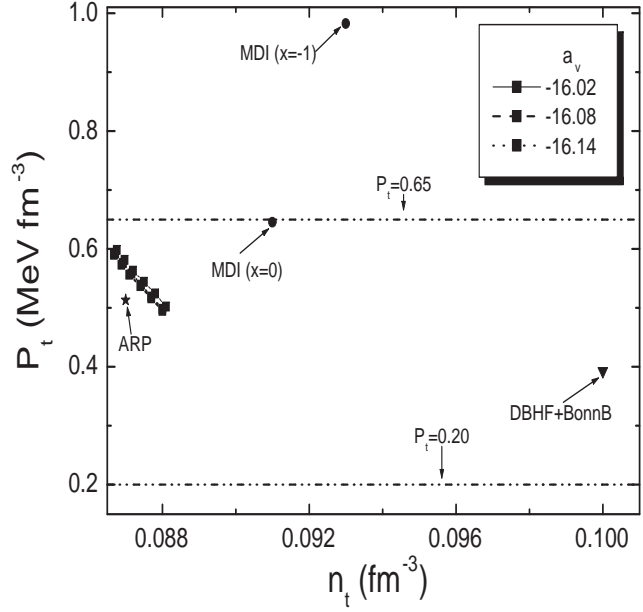}
\caption{The transition pressure $P_t$ as a function of the
transition density $n_t$. For a  fixed value of the symmetry energy
at saturation $a_4=33$ MeV, and three values of
the nuclear matter volume energy coefficient $a_v$,
the parameter $<S_2>$ is varied in the interval between 27.6 MeV and
28.6 MeV. The resulting constraints of $P_t$  and $n_t$ are plotted
 in comparison with results obtained using different models (for details see
Ref.~\cite{Krastev-010}). The lines $P_t=0.2$ MeV fm$^{-3}$ and
$P_t=0.65$ MeV fm$^{-3}$ correspond to the measure of the current
uncertainty in the density dependence of the symmetry energy
\cite{Lattimer-07}. }
\label{Fig_C}
\end{figure}
\begin{figure}
\centering
\includegraphics[height=8.5cm,width=8.0cm]{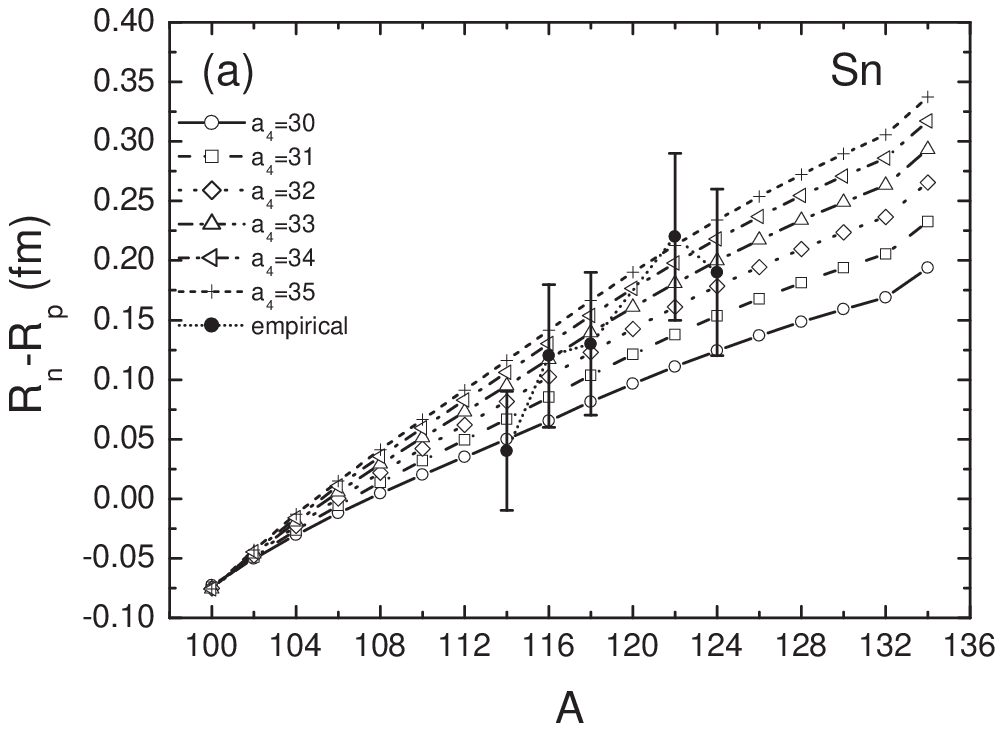}\
\includegraphics[height=8.8cm,width=8.2cm]{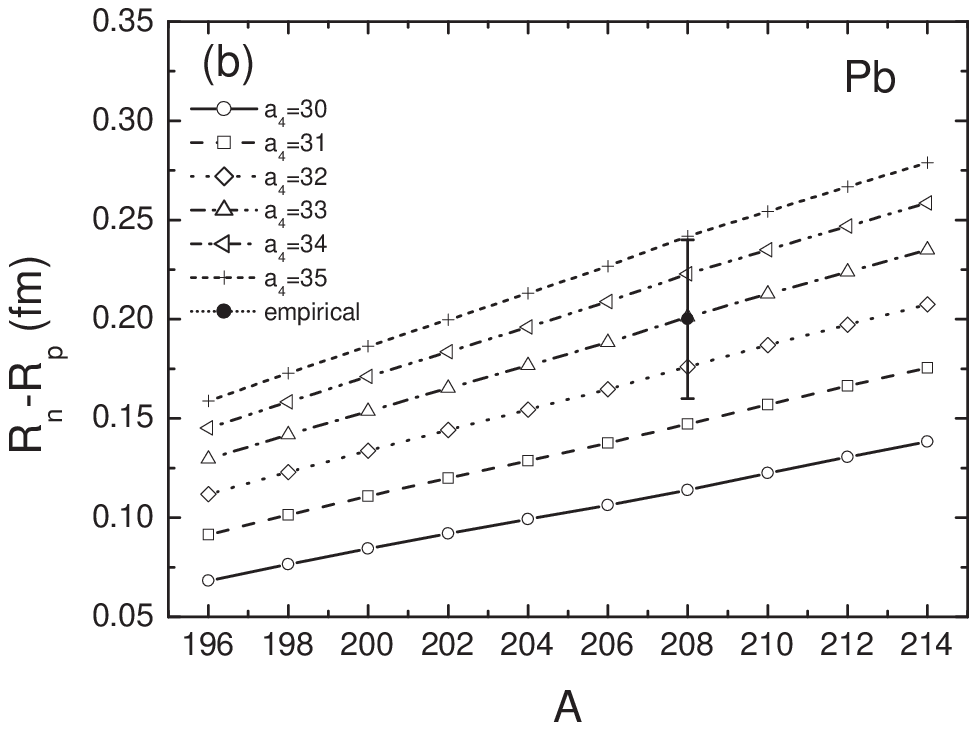}\
\caption{RHB predictions for the differences between neutron and
proton $rms$ radii of (a) Sn and (b) Pb isotopes, in comparison
with available data \cite{Kra.99,SH.94,Kra.94}, for different
choices of the symmetry energy at saturation density $a_4$ (in
MeV).} \label{Fig_D}
\end{figure}
\begin{figure}
\centering
\includegraphics[height=10.0cm,width=10.0cm]{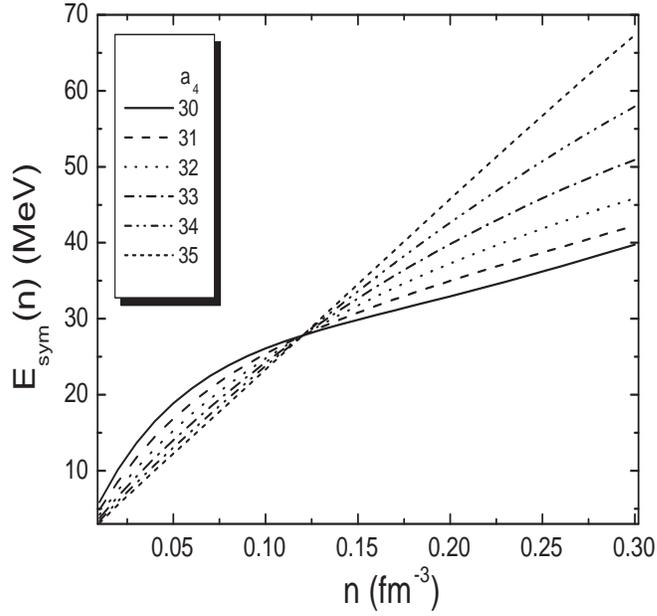}
\caption{The symmetry energy $E_{sym}$ as a function of the
nucleon density $n$ for various values of the symmetry energy at
saturation $a_4$. The parameter $< S_2 > $ (see text) is kept
constant at 27.8 MeV.}
\label{Fig_E}
\end{figure}
\begin{figure}
\centering
\includegraphics[height=8.5cm,width=8.5cm]{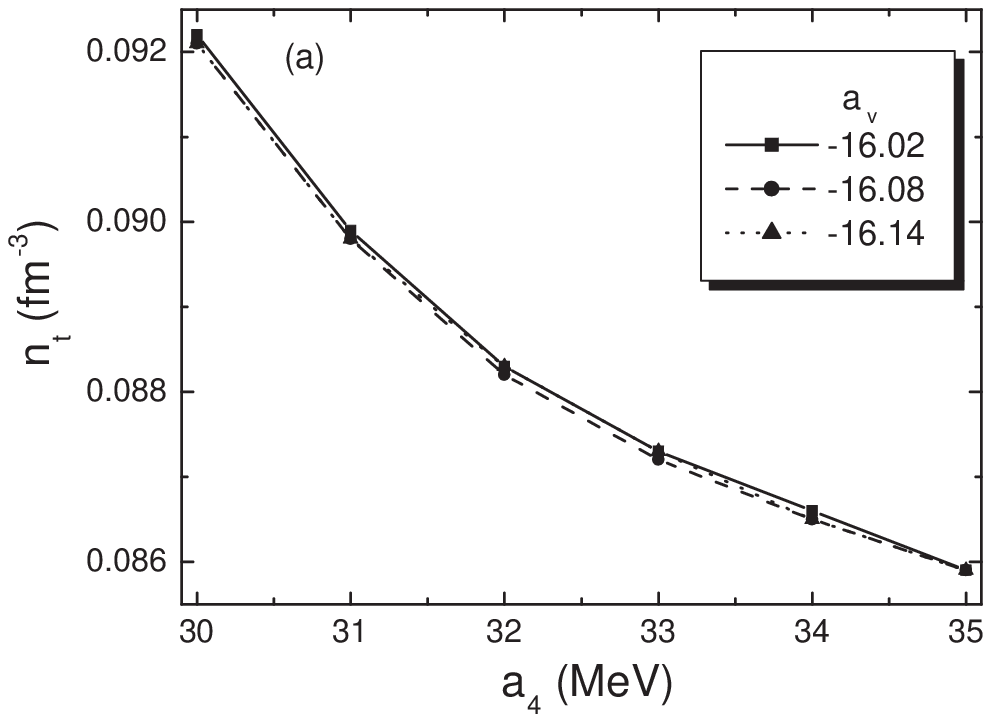}\
\includegraphics[height=8.5cm,width=8.5cm]{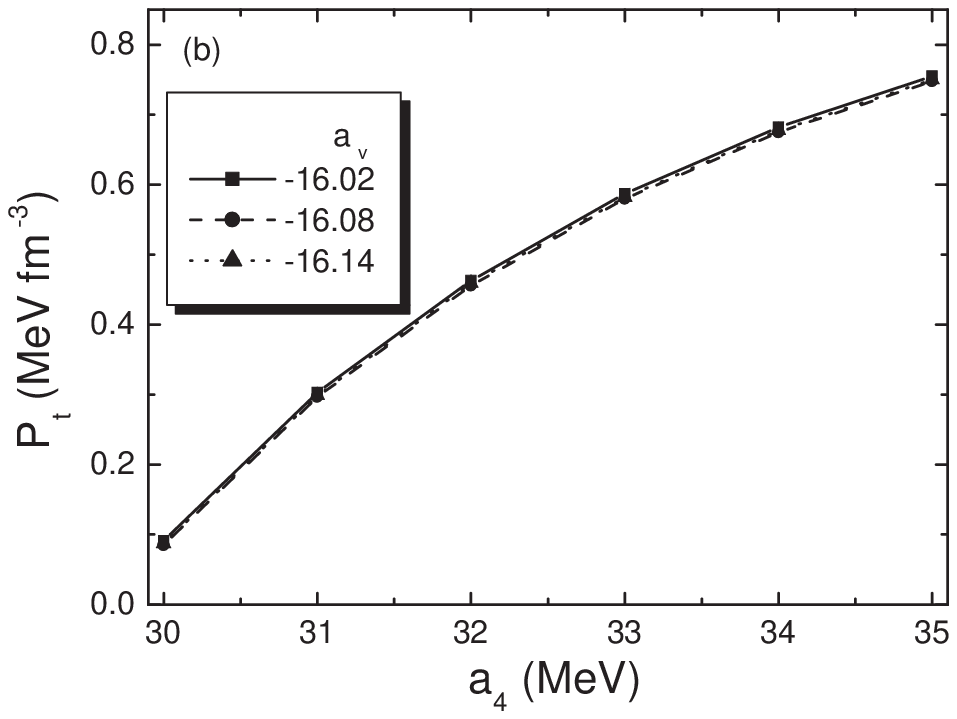}\
\caption{The transition density $n_t$ (a), and the transition
pressure $P_t$ (b), as functions of the symmetry energy at
saturation density $a_4$, for three values of
the nuclear matter volume energy coefficient $a_v$.}
\label{Fig_F}
\end{figure}
\begin{figure}
\centering
\includegraphics[height=10cm,width=10cm]{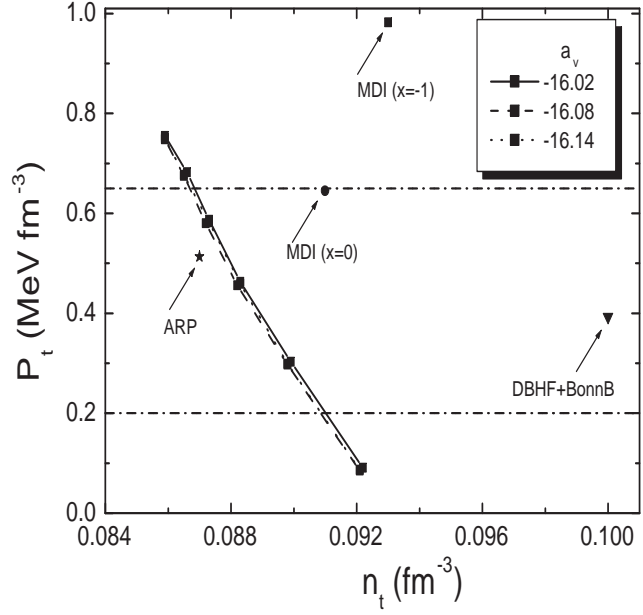}
\caption{Same as described in the caption to Fig.~\ref{Fig_C} but
for fixed $<S_2> = 27.8$ MeV, and the symmetry energy at
saturation in the interval 30 MeV $\leq a_4 \leq$ 35 MeV.}
\label{Fig_G}
\end{figure}
\begin{figure}
\centering
\includegraphics[height=9.5cm,width=7.5cm]{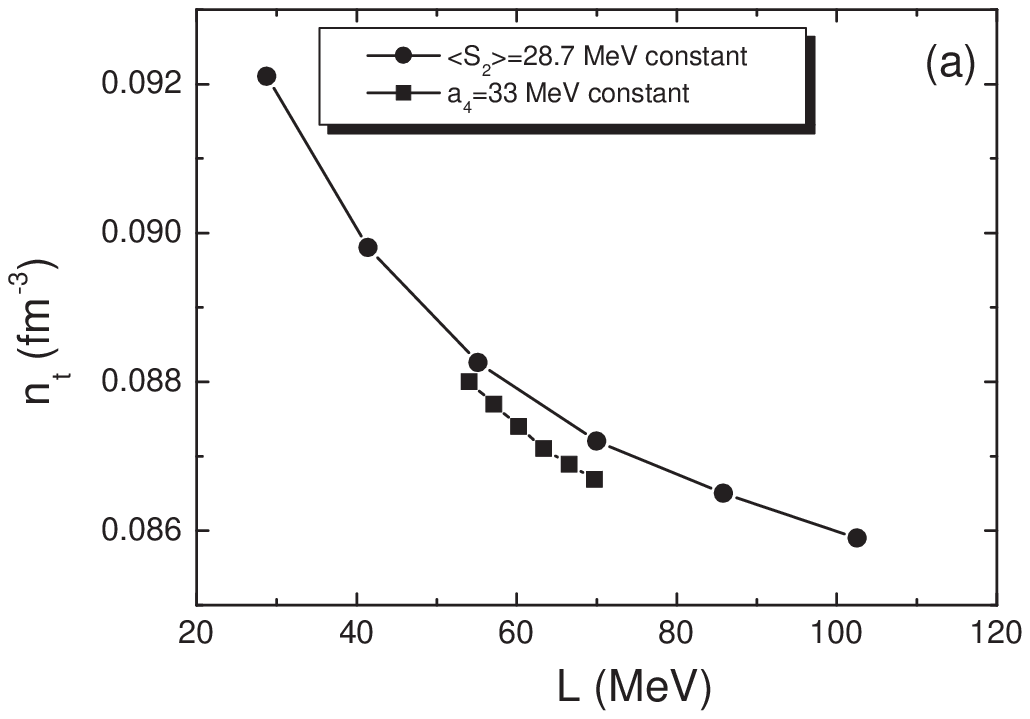}\
\includegraphics[height=9.5cm,width=7.5cm]{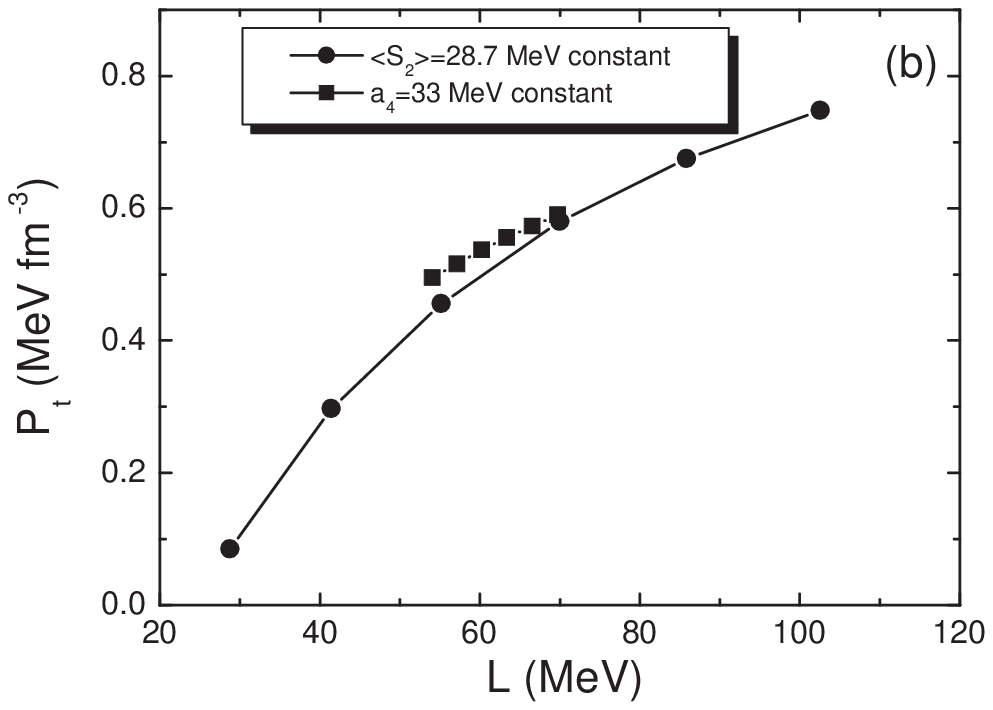}\
\caption{The transition density $n_t$ (a), and the transition
pressure $P_t$ (b), as functions of the symmetry energy at
saturation density $a_4$, for three values of the nuclear matter
volume energy coefficient $a_v$.} \label{Fig_H}
\end{figure}
\begin{figure}
\centering
\includegraphics[height=9.0cm,width=9.0cm]{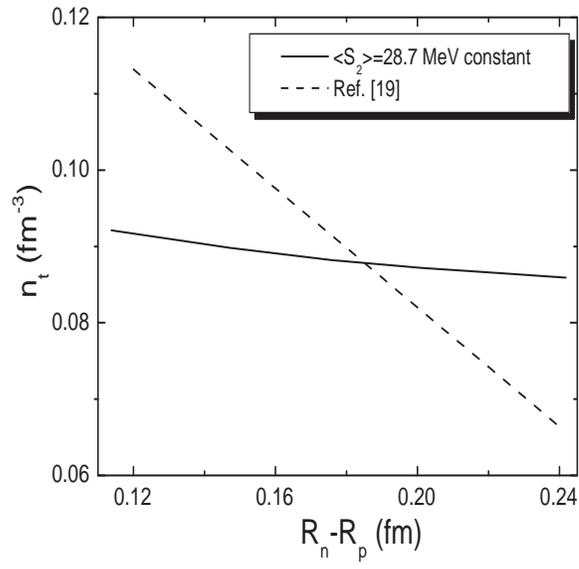}
\caption{The transition density $n_t$ as function of the
neutron-skin thickness $R_n-R_p$ of $^{208}$Pb. The
values of $n_t$ calculated using the thermodynamical model in
the present work (solid), are compared with those of
Ref.~ \cite{Horowitz-01} (see text for description).} \label{Fig_I}
\end{figure}

\end{document}